\documentclass[aps,
                pra,
                twocolumn,
                10pt,
                superscriptaddress,
                longbibliograhy,
                floatfix]{revtex4-2}
\usepackage{hyperref}
\usepackage{amsmath,dsfont}
\usepackage{amsthm}
\usepackage{amssymb}
\usepackage{physics}
\usepackage[dvipsnames]{xcolor}
\usepackage{graphicx}
\usepackage{comment}
\usepackage{mathtools}
\usepackage{enumitem}
\usepackage{tikz}
\usepackage{url}
\usepackage{float}
\usepackage[margin=1in]{geometry}
\usepackage{bbm}
\usepackage{cleveref}
\usepackage{wasysym}
\usepackage{quantikz}
\usepackage{hyperref}
\usepackage{tabularx}
\usepackage{booktabs} 
\usepackage{multirow}
\usepackage{algorithm}
\usepackage[noend]{algpseudocode}
\usepackage{caption,subcaption}
\newlength\tabfsize
\newcolumntype{?}{!{\vrule width 2pt}}
\newlength{\Oldarrayrulewidth}

\newcommand{\eq}[1]{(\ref{eq:#1})}

\newcommand{\calL}{\mathcal{L}}

\definecolor{darkGreen}{RGB}{0,110,0}

\newcommand{\TII}{\affiliation{Quantum Research Center, Technology Innovation Institute, Abu Dhabi, UAE}}

\begin{document}

\title{A competitive NISQ and qubit-efficient solver for the LABS problem}
\author{Marco Sciorilli}
\TII
\affiliation{Laboratoire d’Informatique de Paris 6, CNRS, Sorbonne Universite, 4 Place Jussieu, 75005 Paris, France}
\author{Giancarlo Camilo}
\TII
\author{Thiago O. Maciel}
\TII
\author{Askery Canabarro}
\TII
\author{Lucas Borges}
\TII
\affiliation{Federal University of Rio de Janeiro, Caixa Postal 652, Rio de Janeiro, RJ 21941-972, Brazil}
\author{Leandro Aolita}
\TII

\begin{abstract}
Pauli Correlation Encoding (PCE) is as a qubit-efficient variational approach to combinatorial optimization problems. 
The method offers a polynomial reduction in qubit count and a super-polynomial suppression of barren plateaus.  
Here, we extend the PCE-based framework to solve the Low Autocorrelation Binary Sequences (LABS) problem, a notoriously hard problem often used as a benchmark for classical and quantum solvers. To illustrate this,
we simulate two variants of the PCE quantum solver for LABS instances of up to $N=45$ binary variables: one with commuting and one with maximally non-commuting sets of Pauli operators. 
The simulations use $4$ qubits and a circuit Ansatz with a total of $30$ two-qubit gates. 
We benchmark our method against the state-of-the-art classical solver and other quantum schemes. We observe improved scaling in the total time to reach the exact solution, outperforming the best-performing classical heuristic while using only a fraction of the quantum resources required by other quantum approaches. In addition, we perform proof-of-principle demonstrations on \textrm{IonQ}'s \texttt{Forte} quantum processor, showing that the final solution is resilient to noise.
Our findings point at PCE-based solvers as a promising quantum-inspired classical heuristic for hard problems as well as a tool to reduce the resource requirements for actual quantum algorithms. 
\end{abstract}

\maketitle

The Low Autocorrelation Binary Sequences (LABS) problem ~\cite{Packebusch_2016,CSPLib,zhang2025newimprovementssolvinglarge,Shaydulin_2024,Bernasconi_1987,GALLARDO20091252,BOSKOVIC2017262,Golay1982} is a well-known NP-hard problem with applications in engineering, communications, and statistical physics. 
It offers a challenging benchmark for optimization algorithms~\cite{CSPLib}, as finding the optimal global solutions becomes intractable even for moderate problem sizes $N$. 
Classical methods, though powerful for small instances, face significant challenges in finding good solutions for large $N$, even when employing massively parallel approaches on CPUs and GPUs~\cite{zhang2025newimprovementssolvinglarge}. 
Quantum algorithms have recently been proposed as a potential alternative to classical heuristics for solving combinatorial optimization problems more efficiently~\cite{farhi2014}. 
Recent work showed that the Quantum Approximate Optimization Algorithm (QAOA)~\cite{farhi2014} can solve LABS with a runtime scaling of \( \order{1.46^N} \), which can be further boosted to \( \order{1.21^N} \) if combined with a Grover-like quantum minimum-finding subroutine~\cite{Shaydulin_2024}. 
However, such a quadratic improvement requires a large-scale fault-tolerant quantum computer, 
which is currently not available. 
This letter investigates the effectiveness of solving the LABS problem with NISQ devices using Pauli correlation encoding (PCE), a qubit-efficient technique to encode binary variables into quantum states recently introduced in~\cite{Sciorilli2025}. 
Notably, PCE encodes $N$ binary variables using a polynomially smaller number $n$ of qubits, \emph{i.e.}, $N=\text{poly}(n)$, opening up the possibility of exploring large LABS instances with a number of qubits feasible in near-term devices. 
In particular, using numerical simulations we show that the PCE-based LABS solver has the runtime scaling of $\order{1.330^N}$ for even and $\order{1.326^N}$ for odd instances, which improves over the QAOA scaling and is marginally better even than the state-of-the-art classical heuristics, given by Tabu search~\cite{tabu1}. 
In addition, we do a proof-of-principle deployment on \textrm{IonQ}'s QPU for  instance size $N=120$ (to the best of our knowledge, the largest instance of the problem demonstrated on quantum hardware to date). We observed that training requires only a modest number of shots and that the final solution quality is robust to device errors.

\paragraph*{The LABS problem.} 
Given an integer $N>2$, the LABS problem consists in finding a binary sequence $x \in \{-1,+1\}^{\otimes N}$ of length $N$ that minimizes the \textit{sidelobe energy}
\begin{align}
    E(x) = \sum_{\ell=1}^{N-1} C_\ell^2(x), \quad C_\ell(x) = \sum_{i=1}^{N-\ell} x_i\,x_{i+\ell}\label{eq:compact}\,,
\end{align}
where $C_\ell(x)$ is the $\ell$-th \textit{autocorrelation} of $x$. 
Notice that $C_0(x)=N$ for all $x$, while $(N-\ell)\bmod2\le|C_\ell(x)|\le N-\ell$ for $\ell>0$. 
In statistical physics terms, $E(x)$ can be interpreted as the energy of a system of $N$ spins subject to long-range 2- and 4-body interactions known as the Bernasconi model~\cite{Bernasconi_1987}. 
Equivalently, the problem amounts to maximizing the \textit{merit factor}, 
\begin{align}
  F(x) = \frac{N^2}{2E(x)}, \label{eq:mf}
\end{align}
which is (half) the ratio between $C_0(x)^2$ (the \textit{peak}) and the sum of all the other squared autocorrelations (the \textit{sidelobes}). For this reason, sequences with low autocorrelation provide a large peak-to-sidelobe ratio that is a desirable feature for various engineering applications involving pulse modulation~\cite{Golomb1965}. 

The LABS problem is known to be NP-hard. 
Solutions are degenerate, and the optimization landscape has a peculiar \lq\lq golf course\rq\rq~structure~\cite{Mertens_1998}, plagued by a large number of local minima and extremely isolated global minima~\cite{zhang2025newimprovementssolvinglarge}. 
For small enough $N$, the optimal sequence can be found by brute force enumeration of all $2^N$ possible sequences, but the exponential growth in search space quickly becomes impractical. 
For odd $N$, exact solutions can often (but not always) be found by restricting the search to sequences satisfying the skew-symmetry condition $x_{(N+1)/2+\ell}=(-1)^\ell\,x_{(N+1)/2-\ell}$ for all $\ell\in[1,(N-1)/2]$, which reduces the effective search space size to $2^{N/2}$~\cite{Packebusch_2016}. 
In practice, provably optimal solutions are only known for $N\le66$ 
\cite{Packebusch_2016}. 
For larger instances, different heuristic algorithms with milder exponential scalings but no optimality guarantees 
have been employed~\cite{liu2024search,wang2022low,jia2022computational,shoaib2022hybrid,gancayco2020efficient}. 
The best performing one, Memetic Tabu Search~\cite{tabu1}, has been empirically shown to achieve a runtime scaling of $\order{1.34^N}$ in the range $N\le40$ \cite{GALLARDO20091252,BOSKOVIC2017262,Shaydulin_2024}. 
Although these milder scalings encourage attempting larger instances, for $N > 200$, the solution quality 
of classical heuristics has been observed to degrade significantly, and the asymptotic behavior of the optimal merit factor $F_N\coloneq \max_{x\in\{-1,+1\}^{\otimes N}}F(x)$ for large $N$ remains an open question. 
Golay conjectured based on the ergodicity postulate the asymptotic upper bound $F_N\lesssim12.3248/(8\pi N)^{\frac{3}{2N}}$ as $N\to\infty$~\cite{Golay1982}, but the best known heuristic sequences achieve the much smaller value $F_N\approx6.3421$~\cite{JEDWAB2013882,jedwab2013littlewoodpolynomialssmalll4}. 
Fig.~\ref{fig:labs} in App.~\ref{app:sota} shows a visual summary of the state-of-the-art in LABS solutions using classical solvers. 

\vspace{.5cm}

\paragraph*{Qubit-efficient LABS solver. }
Here we propose a qubit-efficient solver for the LABS problem following the {\it Pauli-correlation encoding} (PCE) framework recently introduced in~\cite{Sciorilli2025}. 
Given a set $\Pi\equiv\{\Pi_i\}_{i\in[N]}$ of $n$-qubit Pauli strings $\Pi_i$, PCE encodes binary variables $x_i$ into correlators from this set as
\begin{align}\label{eq:encoding}
    x_i := \text{sgn}\big(\langle \Pi_i\rangle\big), \quad  i\in[N]\,,
\end{align}
where $\text{sgn}$ is the sign function and $\langle \Pi_i\rangle:=\bra{\Psi}\Pi_i\ket{\Psi}$ denotes the expectation value of $\Pi_i$ on a quantum state $\ket{\Psi}$. 
Since there are $4^n-1$ available traceless Pauli operators, this allows a compressed encoding using $n\ll N$ qubits. 
Below we discuss two explicit choices for the set $\Pi$ used throughout this work. 
We take $\ket{\Psi}$ as the output of a parameterized quantum circuit with a brickwork architecture~\cite{Sciorilli2025} and parameters $\theta$ to be classically optimized using a variational approach (see App.~\ref{app:circuit} for details). 
The goal of the parameter optimization is to minimize the non-linear loss function
\begin{align}\label{eq:loss}
    \calL &= \sum_{\ell=1}^{N-1} \sum_{i,j=1}^{N-\ell} \widetilde{x}_i\,\widetilde{x}_{i+\ell}\,\widetilde{x}_j\,\widetilde{x}_{j+\ell} - \beta\sum_{i=1}^N \widetilde{x}_i^2\,,
\end{align} 
where $\widetilde{x}_i\coloneq\tanh\big(\alpha\,\langle \Pi_i\rangle\big)$ is a real-valued relaxation of the binary variable \eqref{eq:encoding} better-suited for gradient-based optimization. 
The rescaling factor $\alpha>1$ is introduced to restore the non-linear behaviour of the tanh, given that the $n$ qubit correlators have magnitude decreasing polynomially with $n$. 
The last term in \eq{loss} is a regularization term forcing all correlators away from zero, which was observed to improve the solver's performance (details on the choice of regularization and the hyperparameters $\alpha,\beta$ are presented in App.~\ref{app:tuning}). 
Once the training is complete, the circuit output state is measured and a bit-string $x$ is obtained via Eq.~\eq{encoding}. 
In the following we benchmark the performance of PCE with two different choices for the Pauli correlator set $\Pi$.  
The first, $\Pi^{(\text{C})}$, contains randomly selected but mutually \emph{commuting} Paulis. 
The second, $\Pi^{(\text{NC})}$, in contrast, is designed so as to maximize the number of mutually \emph{non-commuting} Pauli operators.  
The intuition is that non-commutativity gives rise to mutually unbiased measurements~\cite{bandyopadhyay2002}, \emph{i.e.}, measurement outcomes that provide as much new information about the underlying state $\ket{\Psi}$ as possible~\cite{wootters1989}, which should increase the independence (hence the expressivity) of the assigned variables $x_i$ in Eq. \eqref{eq:encoding}. 
We refer to App.~\ref{app:noncommuting} for an explicit description of the procedure to construct $\Pi^{(\text{C})}$ and $\Pi^{(\text{NC})}$.

\begin{figure}[t]
    \includegraphics[width=\columnwidth]{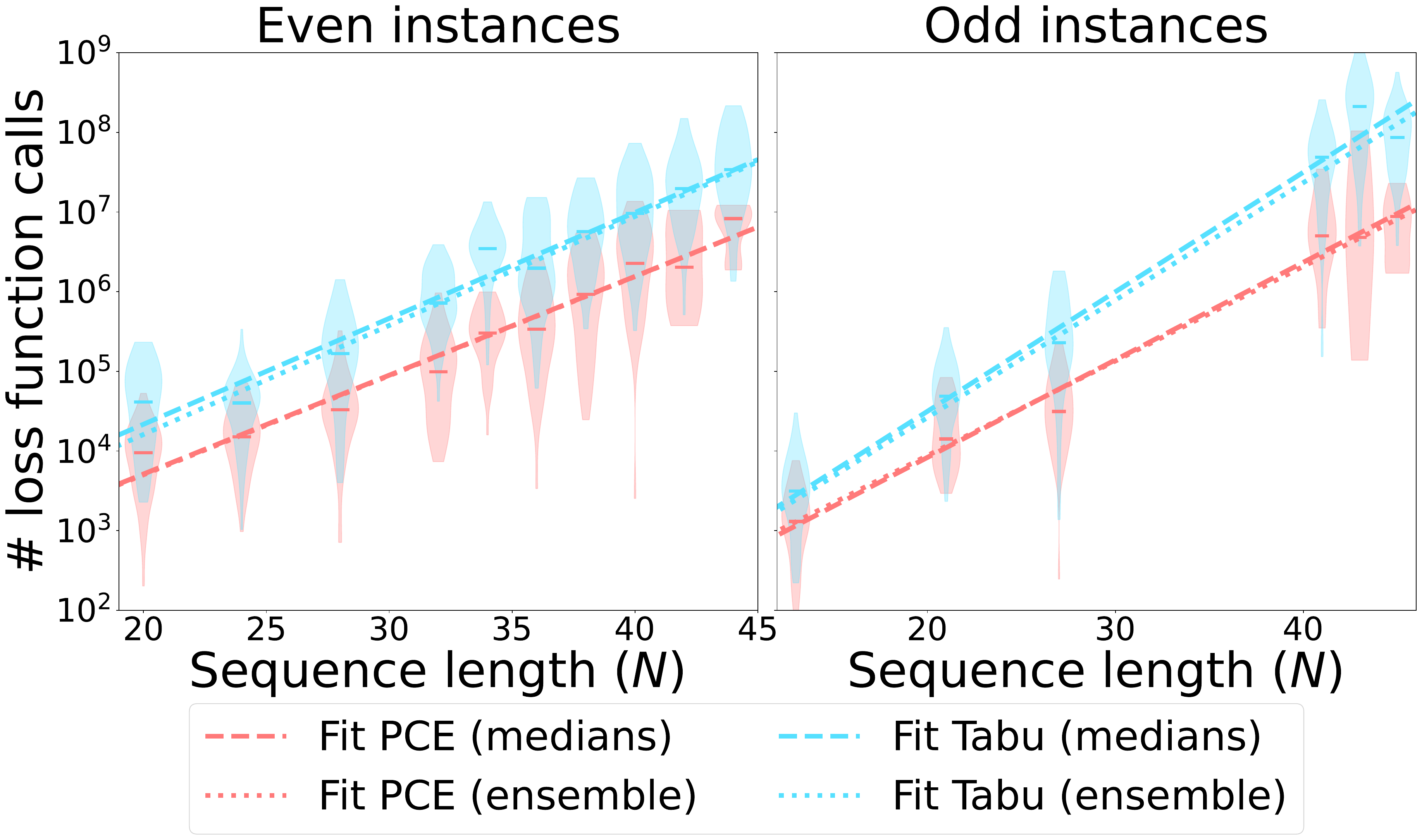}%
    \caption{\textbf{Time-to-solution benchmark.} Scaling comparison of the TTS for the quantum LABS solver based on Pauli Correlation Encoding using $\Pi^{(\text{NC})}$ (PCE, red) \emph{vs.} Tabu Search (Tabu, blue) 
    for even (left) and odd (right) values of $N$. Here, TTS is defined as the total number of cost function evaluations required to reach the exact solution~\cite{BOSKOVIC2017262}. Dashed-dotted lines denote linear regression over the list of median values for each $N$, while dotted curves correspond to linear regression over $50$ points per instance size (except for $N\in[41,45]$, with $10$ runs each). 
    The distribution of points regarding different initializations is depicted by violin plots. 
    The fits show a clear exponential scaling in all the cases, with both median and ensemble statistics 
    consistent with a power-law advantage of PCE over the classical baseline. The precise scaling exponents and confidence intervals are shown in Table~\ref{tab:fits_full_PCE} in App.~\ref{app:approx}.
    }
    \label{fig:tts-bench}
\end{figure}

\vspace{.5cm}

\paragraph*{Numerical results.} 
We perform numerical simulations using the two PCE-based quantum solvers for the LABS instances of sizes $N\le45$ using the Pauli correlation sets $\Pi^{(\text{C})}$ and $\Pi^{(\text{NC})}$ with $n=4$ qubits each.
$\Pi^{(\text{NC})}$ performed consistently better than $\Pi^{(\text{C})}$, so we chose it to benchmark against the other solvers.
We tuned the circuit depth to achieve the best possible approximation ratio as explained in detail in App.~\ref{app:circuit}. 
Notably, the optimal circuit depth scales linearly with $n$ (hence sublinearly with $N$). 
In both cases, we fixed the number of two-qubit gates to $30$, corresponding to $15$ layers of the brickwork circuit and a total of $150$ parameters.

\begin{figure}[t]
    \includegraphics[width=\columnwidth]{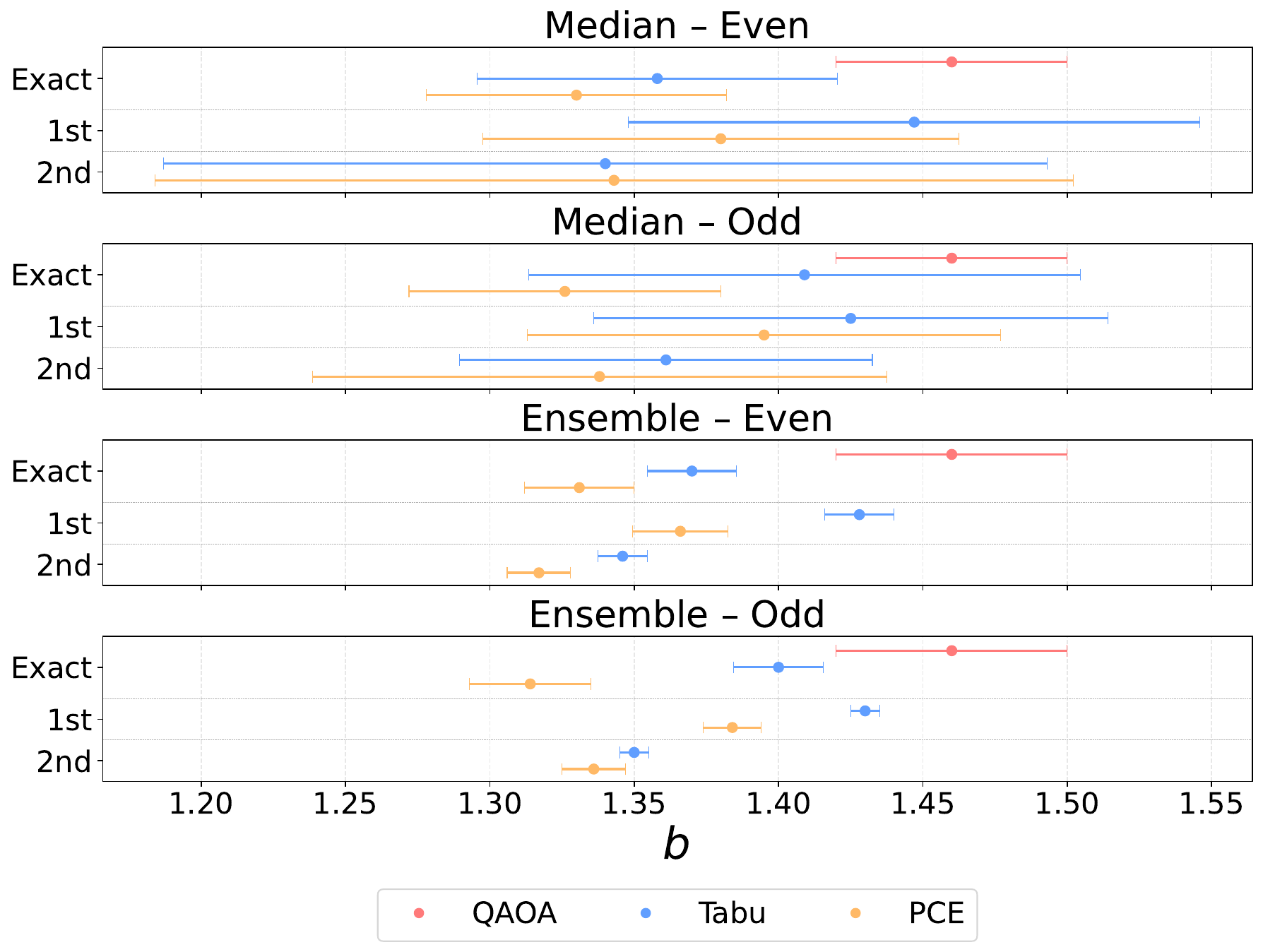}%
    \caption{\textbf{TTS scaling for different solvers.} Fitted basis coefficient $b$ of the exponential scaling in Eq. \eqref{eq:scaling} for our PCE quantum solver using $\Pi^{(\text{NC})}$, QAOA~\cite{Shaydulin_2024}, and Tabu Search (classical) LABS solvers. 
    The corresponding fit data appears in Table~\ref{tab:fits_full_PCE} in App.~\ref{app:approx}.  
    We report the results for TTS, $\text{TTS}_{1\text{st}}$, and $\text{TTS}_{2\text{nd}}$ results, corresponding respectively to the exact solution (Exact) and to approximate solutions given by the first (1st) and second (2nd) excited energy levels.  
    }
    \label{fig:TTS_basis}
\end{figure}

As our figure of merit, we adopt the time-to-solution (TTS), here defined as the number of cost function evaluations required to observe the exact solution, following the approach outlined in~\cite{BOSKOVIC2017262}. 
Moreover, although the time required to reach optimality is a rigorous metric, we assess also the effectiveness of \emph{approximating} LABS solutions by analyzing the time required to reach the second- and third-best objective function values, referred as $\text{TTS}_{1\text{st}}$ and $\text{TTS}_{2\text{nd}}$, respectively. 
For many combinatorial optimization problems, finding a good approximation of the objective function is also hard. 
To obtain our empirical scalings, we used problem instances with unique canonical solutions (\emph{i.e.}, the ground state is only degenerate due to the inherent symmetries of the problem~\cite{BOSKOVIC2017262}): $N\in\{20, 24, 28, 32, 34, 36, 38, 40, 42, 44\}$ for even and $N\in\{13, 21, 27, 41, 43, 45\}$ for odd instances.

\begin{table}[!ht]
    \renewcommand{\arraystretch}{1.3}
    \begin{tabularx}{\columnwidth}{@{\extracolsep{\fill}\hspace{0pt}}|l|c|ccc|}
      \hline
      \textbf{Heuristic} &  $\mathbf{N}$ & \textbf{b} & \textbf{CI} & \textbf{$R^2$} \\
      \hline
      \multicolumn{1}{|l|}{QAOA~\cite{Shaydulin_2024}} & All & 1.46 & 1.42--1.50 & $>$0.94 \\
      \hline
      \multirow{2}{*}{Tabu~\cite{BOSKOVIC2017262,zhang2025newimprovementssolvinglarge}} &  \,Even\, & 1.358 & 1.297--1.422 & 0.97 \\
                                                                                          & Odd  & 1.409 & 1.317--1.508 & 0.97 \\
      \hline
      \multirow{2}{*}{PCE}         & Even & 1.330 & 1.276-1.387 & 0.97 \\
                                   & Odd  &  1.325 & 1.273-1.380 & 0.97 \\
      \hline
    \end{tabularx}%
  \caption{\textbf{TTS fit parameters.} 
  Estimated base coefficient $b$ controlling the exponential scaling of the TTS as in Eq.~\eqref{eq:scaling}, with respective confidence intervals (CI) and $R^2$ values of the linear fit of $\log(\text{TTS})$ for the solvers (PCE, QAOA, Tabu Search) for even and odd instances.}
  \label{tab:labs_results_median}
\end{table}

We benchmark different solvers using the same set of instances, running the optimization $50$ times per instance size $N$ (except for $N\in[41,45]$, with $10$ runs each), separating the even and odd cases since they are expected to have the same scaling, but with different constant factors~\cite{BOSKOVIC2017262} (see Fig.~\ref{fig:tts-bench}). 
We then used this data to fit the exponential curve 
\begin{align}\label{eq:scaling}
    \text{TTS} = c\cdot b^N
\end{align}
via linear regression on $\log(\text{TTS})$ using both the median TTS for each sequence length $N$, and also the full data ensemble (and similarly for $\text{TTS}_{1\text{st}}$ and $\text{TTS}_{2\text{nd}}$). 
The median fit data for the TTS basis coefficient $b$ is presented in Tab.~\ref{tab:labs_results_median}, while the complete fit data (median and ensemble) for TTS, $\text{TTS}_{1\text{st}}$ and $\text{TTS}_{2\text{nd}}$, are detailed in Tab.~\ref{tab:fits_full_PCE} in App.~\ref{app:approx}. 
The results are summarized in Fig.~\ref{fig:TTS_basis}. 

Our results show a clear scaling advantage of PCE compared with the QAOA quantum solver of~\cite{Shaydulin_2024} for both even and odd instances. 
For a fair comparison, the final quantum amplitude amplification step suggested in~\cite{Shaydulin_2024} was omitted here, since takes the solver beyond the scope of near-term devices. 
In addition, we used Tabu without its memetic component, which was observed to not offer any scaling advantage~\cite{BOSKOVIC2017262}, and tenure parameter chosen from the recent GPU implementation in~\cite{zhang2025newimprovementssolvinglarge}. 
All the TTS scalings were obtained by local simulations. 
The PCE results are competitive (within error bars), and in the case of approximate solutions even consistently better, with respect to Tabu search, considered the state-of-the-art classical heuristics for LABS~\cite{BOSKOVIC2017262}, revealing a separation that exceeds the error bars, see App.~\ref{app:approx}.
While minor polynomial advantage factors do not alter the overall exponential scaling, in practice even minor improvements in the basis might play an important role when considering larger problem instances~\cite{Sciorilli2025} (see details in App.~\ref{app:crossover}). 
Thus, the approach is promising even as a quantum-inspired algorithm, although in that case one must factor in the significant overhead from classical simulation of the quantum circuit.
Finally, assuming the observed TTS scalings remain the same for larger instances, to run PCE in a quantum hardware, we still need to account the overhead of the gate operations of running quantum circuits multiple times to estimate the loss function and its gradients to a given precision $\epsilon$. The cost of gate operations are platform-specific, but assuming all gates done sequentially, it scales as $\mathcal{O}(N)$ (see App.~\ref{app:circuit}); the sample complexity is $\mathcal{O}(N^7/\epsilon^2)$ (see App.~\ref{app:sample}); and to estimate the gradient we need $\mathcal{O}(N)$ calls. Thus, the quantum approach incurs an overall overhead of $\mathcal{O}(N^9/\epsilon^2)$.
Even considering the overall overhead of $\mathcal{O}(N^9)$ when running PCE in a quantum hardware, it is still possible to obtain a preliminary estimate of a crossover point where quantum advantage might be achieved. In App.~\ref{app:crossover}, we showed that, from instance sizes of order of thousands, we would expect advantage in terms of query access to the exact solution, but since the classical solvers can not output exact solutions on instances larger than $N>66$, we might see better (approximate) solutions from the quantum approach sooner.

\vspace{.3cm}

\paragraph*{Experimental results.} 
In addition, we perform a proof-of-principle experimental demonstration on the \textrm{IonQ}'s \texttt{Forte} quantum processor. 
For that, instead of using the randomly selected sets of Paulis used in the numerical demonstrations, here we used the same set used in~\cite{Sciorilli2025} of $k$-body correlators on $n$ qubits of the form $X^k\openone^{n-k}, Y^k\openone^{n-k}, Z^k\openone^{n-k}$ for a fixed $k$ (here $n=10$ and $k=2$, which corresponds to a quadratic compression in the number of qubits). 
From the experimental perspective, this is more convenient than $\Pi^\text{(NC)}$ since it requires only $3$ measurement setups.
We have deployed a classically pre-trained circuit on the $N=120$ LABS instance and compared the effect of hardware noise on the dispersion of the loss function (with respect to different initializations) for the first and last optimization steps. 
The full details are presented in App. ~\ref{app:experiment}. 
The results show that shot noise has only a minor impact on solution quality, \emph{i.e.}, a few thousand shots already yield sufficient statistics for experimental training. Hardware noise however, while not preventing training, significantly degrades the final solutions and can undermine the expected TTS scaling advantage. This makes error mitigation, or higher compression rate, a key lever for potential training on near-term QPUs.

\vspace{.3cm}

\paragraph*{Concluding remarks.} 

Two highlights of our approach are its qubit-efficiency and the strong performance obtained on a hard problem like LABS. 
With only $4$ qubits and a total of $30$ two-qubit gates, our solver attains a competitive run-time scaling on a hard sequence-length regime. 
Our results show that the performance of PCE is improved through a careful choice of the Pauli correlation set $\Pi$ so as to maximize the expressivity of the binary variables encoded via Eq.~\eqref{eq:encoding}.
Even though PCE-based variational methods have no performance guarantees, the fact that they consistently display such a strong performance on hard problems such as MaxCut and LABS urges for further investigations.
In addition, we deployed our solver on a commercially available QPU for a large size $N=120$ instance, showing that training can be done with modest shot budgets,
and with a final solution that is robust against hardware noise. Numerical simulations suggest that incorporating quantum error mitigation could make full training possible on near-term devices. 
Moreover, we notice that PCE can also be used as a warm start for classical solvers \cite{cadavid2025}: in App. \ref{app:kike}, we show preliminary numerical explorations indicating that the runtime scaling can be further decreased by running classical solvers on top of the PCE solution.
We note that,for the hardware-friendly variant used in the experiment, $n=8$ qubits would be enough to tackle the largest sequence lengths for which LABS has known exact solutions ($N=66$), while $n\le20$ qubits are enough to challenge even the largest instances solved by state-of-the-art classical heuristics~\cite{P_eni_nik_2025}.
 This regime offers an interesting testbed for our solver as a quantum-inspired heuristic. 
A challenge there would be the high number of loss-function evaluations required. This could be mitigated by incorporating problem symmetries into the circuit ansatz and combining this with exact geodesic transport for parameter updates~\cite{Ferreira-Martins2025}.
Other research directions may be problems beyond LABS, such as those in the Quantum Optimization Benchmark Library (QOBLIB)~\cite{opt_problems}. Our results raise the question of whether PCE has a potential quantum advantage. 
Although the constant overhead of simulating a full quantum state vector may delay the crossover point after which a runtime advantage becomes evident to larger instance sizes, our method still shows promising performance even in a quantum-inspired setting. All these are exciting open questions for future work.

\paragraph*{Acknowledgments.} We thank Allan Tosta for fruitful discussions.

\bibliography{refs.bib}

\appendix

\section*{Supplementary Information}

\subsection{Classical state-of-the-art for LABS}\label{app:sota}

\begin{figure}[ht]
    \includegraphics[width=\columnwidth]{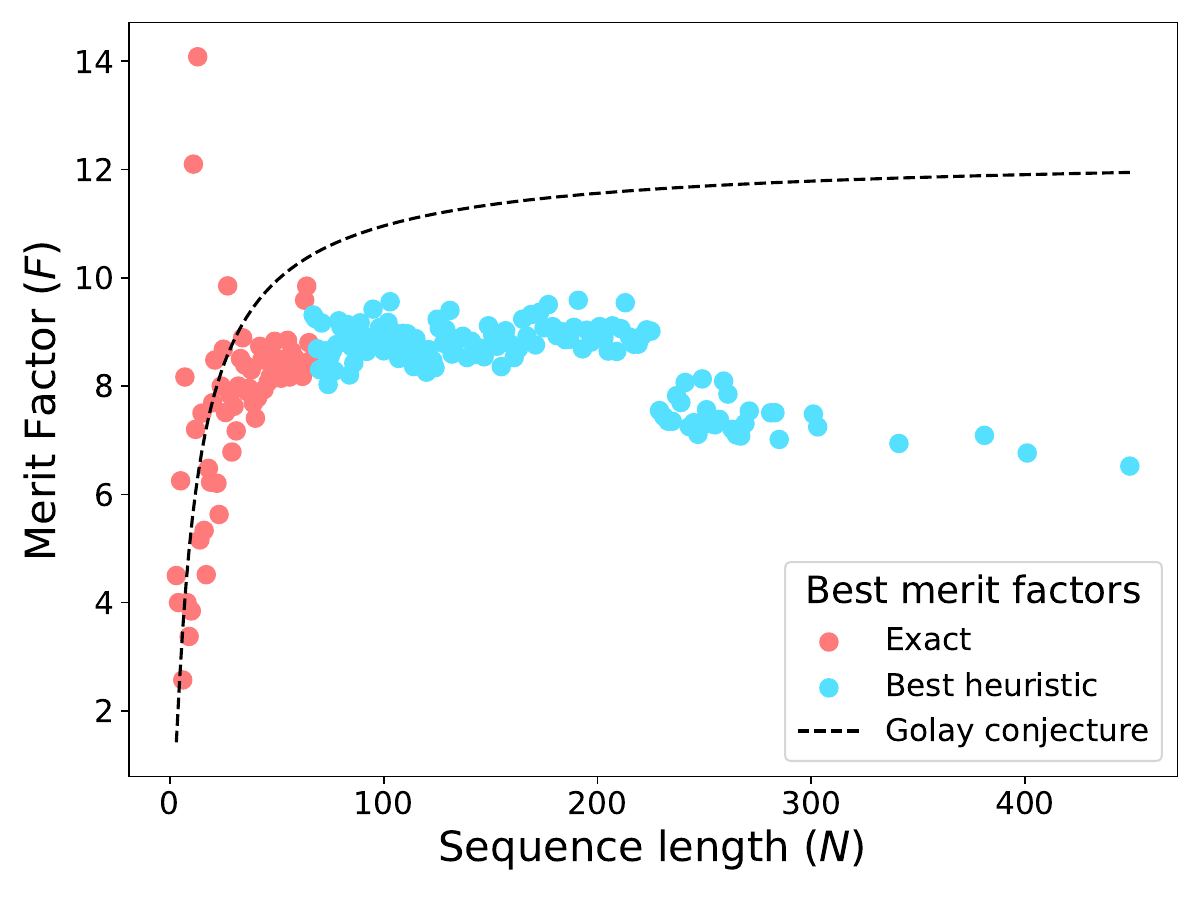}
    \caption{{\bf LABS state-of-the-art.} Best merit factor vs. problem size $N$ for classical solvers. Red dots indicate the optimal merit factor $F_N\coloneq \max_{x\in\{-1,+1\}^{\otimes N}}F(x)$ obtained via exact solvers, while blue dots indicate the best known approximation to $F_N$ obtained via heuristic solvers. 
    The dashed curve is Golay's conjectured asymptotic upper bound $\approx12.3248/(8\pi N)^{\frac{3}{2N}}$~\cite{Golay1982}.}
    \label{fig:labs}
\end{figure}

Here we summarize the state-of-the-art of solutions to the LABS problem obtained by classical solvers. 
We use the data are publicly available at~\cite{LABS_github}, combined with recent improvements in particular instances reported in~\cite{zhang2025newimprovementssolvinglarge}. 
For $N\le66$, solutions with the optimal solution merit factor 
\begin{align}
    F_N\coloneq \max_{x\in\{-1,+1\}^{\otimes N}}F(x)
\end{align}
can be found using exhaustive search or branch and bound methods. 
For larger $N$, only heuristic solutions are known and there is no proof of optimality. 
The best known solutions were obtained using the Memetic Tabu Search meta-heuristics~\cite{BOSKOVIC2017262}. 
Fig.~\ref{fig:labs} depicts the best known solutions for $2\le N\le449$. 
The largest known merit factor is achieved for $N=13$ (namely $F_{13}=169/12\approx14.083$) by a so-called \emph{Barker sequence}~\cite{barker1953group}. 

\subsection{Circuit complexity} \label{app:circuit}

Here we provide details on the variational circuit ansatz, the classical optimizer, and strategy to fix the circuit depth for the PCE quantum solver.

\paragraph*{Variational circuit.} 
For the variational quantum circuit, we employ a hardware-efficient ansatz with the same gate configuration described in~\cite{Sciorilli2025}, namely, RX, RY, and RZ rotations as single-qubit gates and the Mølmer–Sørensen (MS) gate -- the native entangling gate of trapped-ion quantum computing platforms, which feature currently the highest fidelities among NISQ devices -- for two-qubit interactions. 

\paragraph*{Classical optimizer.} 
For the classical parameter optimization, we used the SLSQP algorithm~\cite{slsqp}, although analogous performance was observed using the Adam optimizer~\cite{adam}.

\begin{figure}[!t]
    \includegraphics[width=\columnwidth]{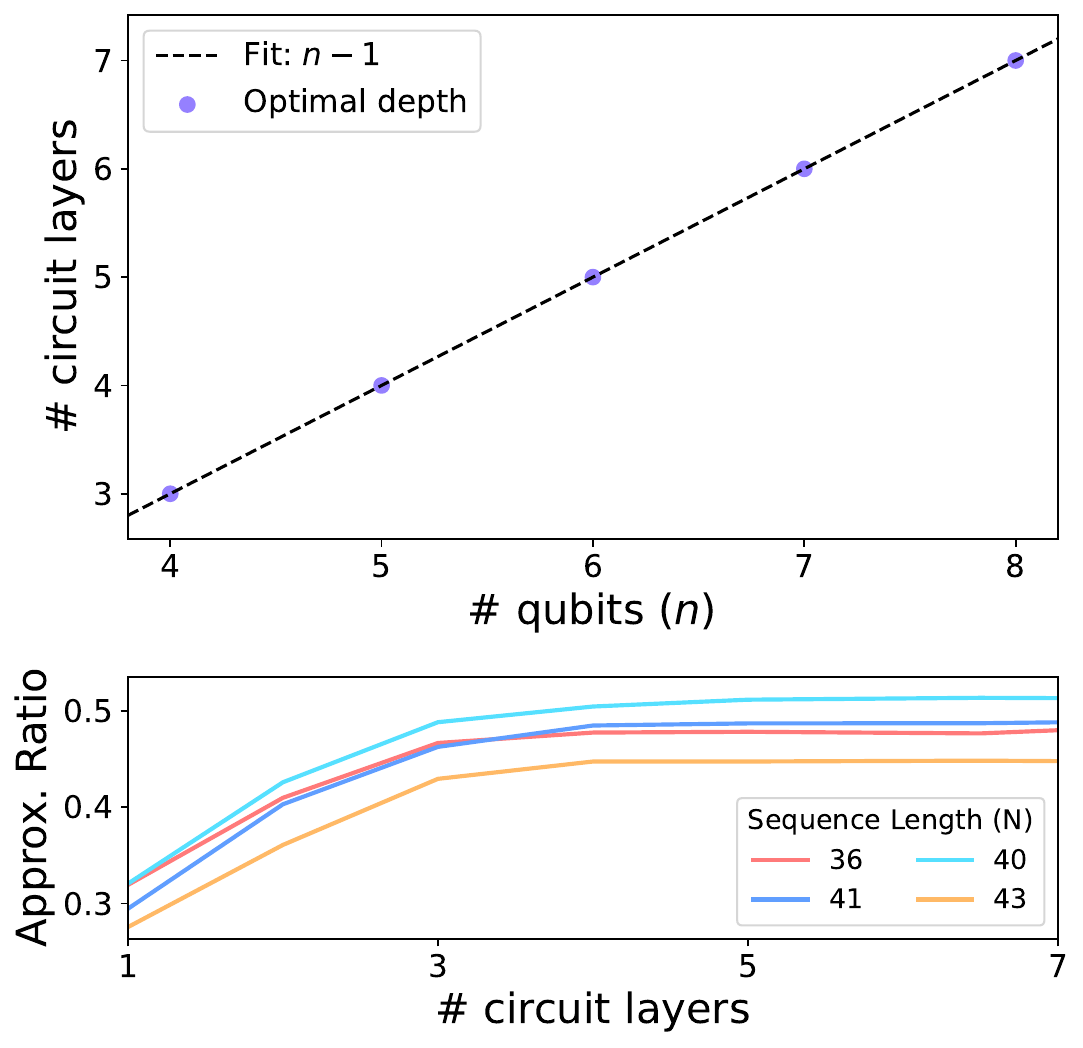}%
    \caption{\textbf{Circuit depth scaling.}  
    \emph{(Top)} Linear scaling of the optimal circuit depth with the number of qubits $n$ for the qubit-efficient solver with quadratic compression using $\Pi^{(\text{C})}$. 
    This corresponds to a $\mathcal{O}(\sqrt{N})$ scaling with the problem size $N$. 
    \emph{(Bottom)}
    For each $n$ (here illustrated for $n=4$), the optimal circuit depth was determined by increasing the number of circuit layers until no significant improvement in the average approximation ratio was observed over $1000$ runs. 
    }
    \label{fig:depth_scaling}
\end{figure}

\paragraph*{Circuit depth scaling.} 
To determine a suitable circuit depth, we chose per each circuit size $4$ different  hard instances $N$ (namely, $2$ even and $2$ odd) as mentioned in the main text and solved them using the PCE quantum solver with a circuit ansatz having an increasing number of layers. 
In each instance and number of layers, we run the solver with $1000$ different random initializations and, for each run, record the corresponding fraction of the optimal merit factor $F_N$ achieved (\emph{i.e.}, approximation ratio). 
Then we used the Kolmogorov–Smirnov test~\cite{hollander2013nonparametric} to identify the point after which there were no more statistically significant improvements in the distributions of the approximation ratios, which we refer to as optimal circuit depth. 
As shown in Fig.~\ref{fig:depth_scaling}, the optimal depth was observed to scale linearly with the number of qubits $n$, and therefore sublinearly with the instance size $N$. 
Interestingly, a linear scaling of the depth in $n$ is also known to lead to approximate Haar randomness~\cite{McClean_2018}. 

\subsection{Regularization term and hyperparameters}\label{app:tuning}

In this section, we discuss the form of the regularization term and the tuning of hyperparameters $\alpha$ and $\beta$ appearing in the loss function \eqref{eq:loss}.

\paragraph*{Regularization term.} 
We first notice from Eq. \eqref{eq:compact} that the energy spectrum of LABS is strictly positive. 
In our formulation of the problem, however, the binary variables are relaxed to (hyperbolic tangents of) expectation values of Pauli operators, which take real values in the interval $[-1,1]$. 
\begin{figure}[ht]
    \includegraphics[width=\columnwidth]{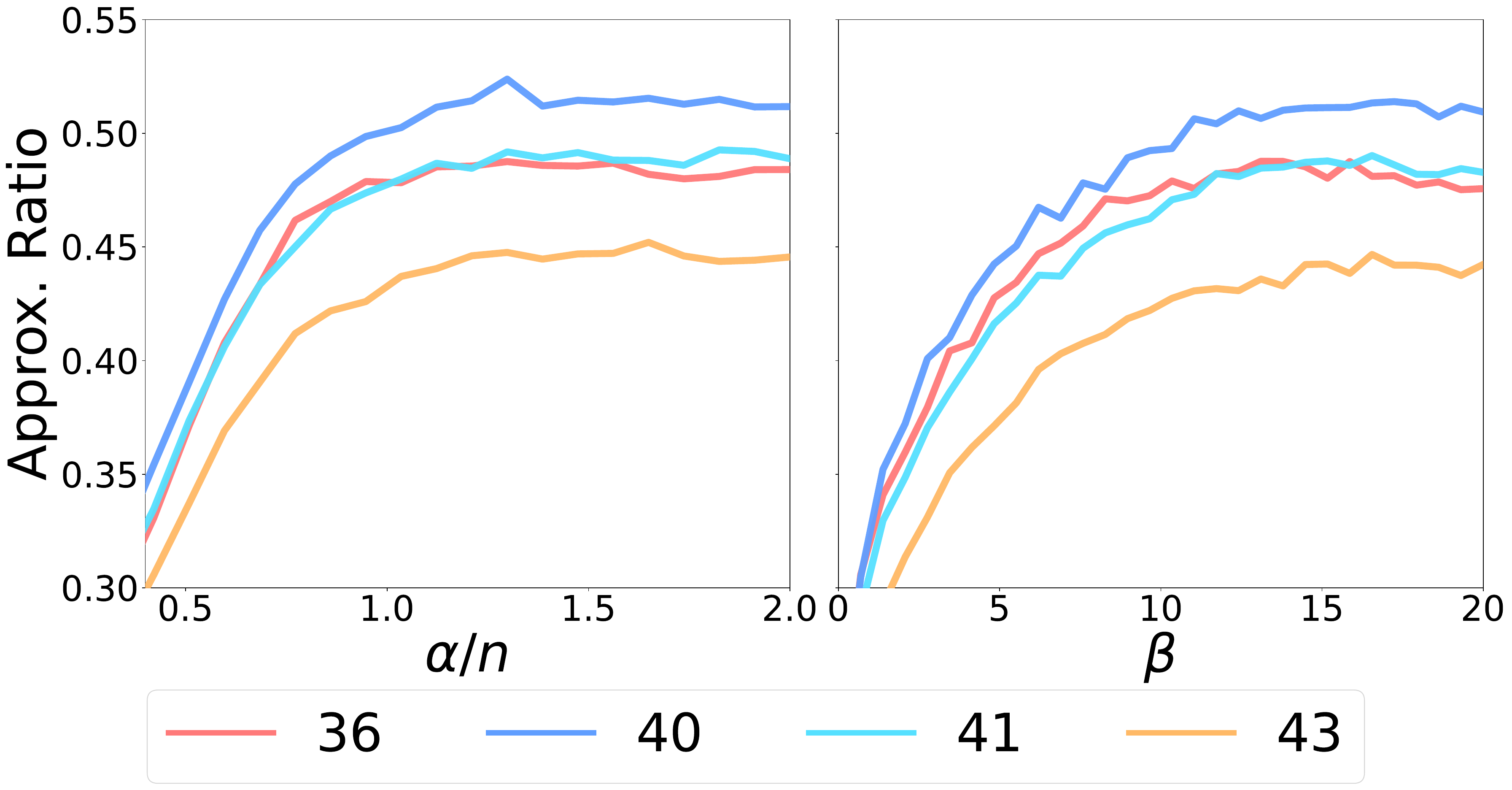}
    \caption{{\bf Hyperparameters tuning.} 
    (\emph{Left}) Tuning of the rescaling parameter $\alpha$;
    (\emph{Right}) Tuning of the penalization constant $\beta$. 
    In each case, we fixed the number of qubits to $n=4$, the circuit depth to $10$, and the optimal value was determined by increasing the corresponding parameter until no significant improvement in the average approximation ratio was observed over $1000$ runs. 
    }
    \label{fig:hyperparams}
\end{figure}
Therefore, if one tries to minimize the loss function~\ref{eq:loss} naively without any penalization term ($\beta=0$), we quickly converge to the trivial solution $\widetilde{x}_i=0$ for all $i$ (where all the expectation values go to $0$), which does not offer any valuable insight for the binary version of the problem. 
For this reason, we introduced a regularization term with a positive coefficient $\beta$ penalizing this strong attractor. For simplicity, we used the simple $\ell_2$-norm penalization $\beta\sum_{i=1}^N \widetilde{x}_i^2$.

\paragraph*{Tuning of $\alpha$ and $\beta$.} 
For each instance and hyperparameter size, we ran the solver using $1000$ distinct random initializations. For every run, we recorded the achieved fraction of the optimal merit factor $F_N$, referred to as the \emph{approximation ratio}, as reported in Fig.~\ref{fig:hyperparams}. Next, we used the Kolmogorov–Smirnov test~\cite{hollander2013nonparametric} to identify the hyperparameter size beyond which further increases did not result in statistically significant improvements in the approximation ratio distributions. As shown in Fig.~\ref{fig:hyperparams} for the case of $n=4$ qubits, this resulted in values $\alpha = 1.5\,n$ and $\beta = 15$. For the purpose of our estimation of the Time-To-Solution (TTS), these were the values used through our numerical benchmarks.

\subsection{PCE as a LABS approximate solver}
\label{app:approx}

 For many combinatorial optimization problems, finding a good approximation of the objective function is also hard. Unlike the MaxCut problem, a rigorous proof of the approximation hardness for the LABS problem is still lacking. However, numerical studies and its unique optimization landscape strongly suggest that LABS is likely difficult to approximate (see~\cite{ferreira2000}).

Even though the time required to reach optimality is a rigorous metric, it does not fully cover an algorithm's overall performance. 
\begin{figure}[!ht]
    \includegraphics[width=\columnwidth]{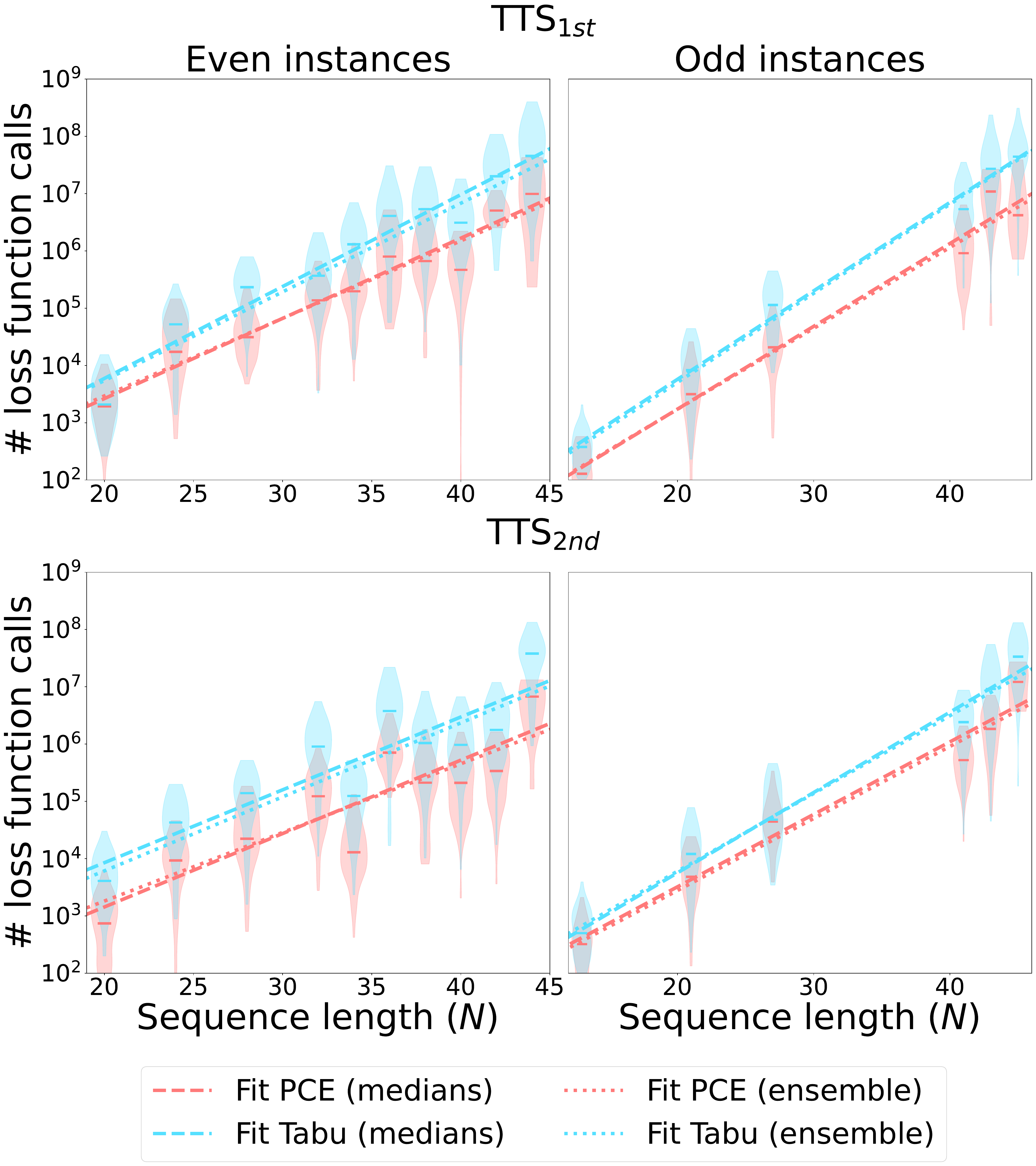}%
    \caption{\textbf{Time-to-solution benchmark.} Scaling comparison of the $\text{TTS}_{1\text{st}}$ and $\text{TTS}_{2\text{nd}}$ for the quantum LABS solver based on Pauli Correlation Encoding (PCE, red) \emph{vs.} Tabu Search (Tabu, blue) 
    for even (left) and odd (right) values of $N$. Here, TTS is defined as the total number of cost function evaluations required to reach the exact solution~\cite{BOSKOVIC2017262}. Dashed-dotted lines denote linear regression over the list of median values for each $N$, while dotted curves correspond to linear regression over the full dataset of $50$ points per instance size. 
    The distribution of points over different initializations is depicted by violin plots at each $N$. 
    The fits show a clear exponential scaling in all the cases, with both median and ensemble statistics 
    consistent with a power-law advantage of PCE over the classical baseline. The precise scaling exponents and confidence intervals are shown in Tab.~\ref{tab:fits_full_PCE}. }
    \label{fig:excited_states}
\end{figure}

Notwithstanding, we evaluate our solver's performance in approximating solutions to the LABS problem by analyzing the time-to-solution (TTS) required to reach the second- and third-best objective function values (\emph{i.e.}, the first- and second-excited energy levels of Eq.~\ref{eq:compact}) to assess approximate solutions to the LABS problem, here referred as $\text{TTS}_{1\text{st}}$ and $\text{TTS}_{2\text{nd}}$, respectively.

In both estimates, we observed an exponential behavior similar to the exact solution. However, the advantage over Tabu Search revealed a separation that exceeds the error bars, see Fig.~\ref{fig:excited_states}, with the corresponding exponential bases reported in Tab.~\ref{tab:fits_full_PCE}. We were unable to obtain the corresponding data for QAOA due to the highly demanding nature of its classical simulations at this scale~\cite{Shaydulin_2024}.

\begin{table*}[htb]
    \renewcommand{\arraystretch}{1.3}
    \begin{tabularx}{.8\textwidth}{@{\extracolsep{\fill}\hspace{0pt}}|c|c|c|c|cc|cc|c|}
    \hline
      \textbf{Solution}& \textbf{Heuristic} & \textbf{Type} & $\mathbf{N}$ & \textbf{b} & \textbf{CI} & \textbf{c} & \textbf{CI} & \textbf{$R^2$} \\
      \hline
      \multirow{11}{*}{Exact} & \multicolumn{1}{l|}{QAOA \cite{Shaydulin_2024}} & All & All & 1.46 & 1.42--1.50 & -- & -- & $>$0.94 \\
      \cline{2-9}
                    & \multirow{4}{*}{Tabu} & \multirow{2}{*}{Median} & \,Even\, & 1.358 & 1.297--1.422 & 47.57 & 9.738--232.4 &\phantom{$>$}0.97 \\
      \cline{4-9}
                      &             &                         & Odd  & 1.413 & 1.334--1.496 & 31.42 & 4.539--217.5 & \phantom{$>$}0.97 \\
      \cline{3-9}
                     &              & \multirow{2}{*}{Ensemble} &  Even & 1.370 & 1.355--1.386 & 29.54 & 19.76--44.16  &\phantom{$>$}0.78 \\
      \cline{4-9}
                  &                 &                         & Odd  & 1.404 & 1.392--1.417 & 29.67  & 21.96--40.09 &\phantom{$>$}0.88 \\
      \cline{2-9}
        & \multirow{4}{*}{PCE}         & \multirow{2}{*}{Median} & Even & \textbf{1.331} & 1.276--1.387 & 16.99 & 4.010--71.98 &\phantom{$>$}0.97 \\
      \cline{4-9}
      &                             &                         & Odd  & \textbf{1.325} & 1.273--1.380 & 29.54  & 7.470--116.8 &\phantom{$>$}0.99 \\
      \cline{3-9}
     &                              & \multirow{2}{*}{Ensemble} &  Even & \textbf{1.332} & 1.313--1.351 & 16.36 & 10.05--26.65 & \phantom{$>$}0.77 \\
      \cline{4-9}
      &                             &                         &  Odd  & \textbf{1.314} & 1.293--1.335 & 36.96  & 21.46--63.64 & \phantom{$>$}0.85 \\
      \cline{2-9}
      & \multirow{2}{*}{PCE + Tabu} & \multirow{2}{*}{Median} & Even & 1.244 &  1.233--1.255 & 2288.2 & 1767.0--3030.2 & \phantom{$>$}0.98 \\
      \cline{4-9}
      &                              &                         & Odd  & 1.330 & 1.323--1.337 & 500.3 & 428.1--591.7 & \phantom{$>$}0.98 \\
      \hline
       \multirow{8}{*}{1st} & \multirow{4}{*}{Tabu} & \multirow{2}{*}{Median} & \,Even\, & 1.447 & 1.351--1.549 & 3.684 & 0.3434--39.53 &\phantom{$>$}0.95 \\
      \cline{4-9}
                      &             &                         & Odd  & 1.427 & 1.369--1.487 & 4.652 & 1.146--18.89 & \phantom{$>$}0.99 \\
      \cline{3-9}
                     &              & \multirow{2}{*}{Ensemble} & Even & 1.428 & 1.416--1.440 & 4.383 & 3.279--5.858 & \phantom{$>$}0.77 \\
      \cline{4-9}
                  &                 &                         & Odd  & 1.431 & 1.426--1.435 & 3.923 & 3.522--4.371 & \phantom{$>$}0.98 \\
      \cline{2-9}
        & \multirow{4}{*}{PCE}         & \multirow{2}{*}{Median} & Even & \textbf{1.395} & 1.301--1.465 & 2.217 & 0.5327--32.91 & \phantom{$>$}0.95 \\
      \cline{4-9}
      &                             &                         & Odd  & \textbf{1.374} & 1.301--1.465 & 2.935  & 0.5328--32.91 & \phantom{$>$}0.98 \\
      \cline{3-9}
     &                              & \multirow{2}{*}{Ensemble} &  Even & \textbf{1.366} & 1.349--1.382 & 5.795 & 3.834--8.759 & \phantom{$>$}0.73 \\
      \cline{4-9}
      &                             &                         &  Odd  & \textbf{1.384} & 1.375--1.394 & 2.559  & 2.005--3.267 & \phantom{$>$}0.89 \\
      \hline
       \multirow{8}{*}{2nd} & \multirow{4}{*}{Tabu} & \multirow{2}{*}{Median} & \,Even\, & \textbf{1.340} & 1.196--1.502 & 24.32 & 0.4693--1260 & \phantom{$>$}0.81 \\
      \cline{4-9}
                      &             &                         & Odd  & 1.381 & 1.314--1.452 & 8.855 & 1.621--48.37 & \phantom{$>$}0.99 \\
      \cline{3-9}
                     &              & \multirow{2}{*}{Ensemble} & Even & 1.346 & 1.338--1.355 & 16.03 & 12.84--20.00  & \phantom{$>$}0.65 \\
      \cline{4-9}
                  &                 &                         & Odd  & 1.367 & 1.363--1.371 & 11.61 & 10.42--12.93 & \phantom{$>$}0.91 \\
      \cline{2-9}
        & \multirow{4}{*}{PCE}         & \multirow{2}{*}{Median} & Even & 1.343 & 1.193--1.512 & 3.923 & 0.06551--235.0 & \phantom{$>$}0.81 \\
      \cline{4-9}
      &                             &                         & Odd  & \textbf{1.338} & 1.243--1.441 & 9.486  & 0.7658--117.5 & \phantom{$>$}0.97 \\
      \cline{3-9}
     &                              & \multirow{2}{*}{Ensemble} & Even & \textbf{1.318} & 1.306--1.329 & 7.319 & 5.450--9.829 & \phantom{$>$}0.61 \\
      \cline{4-9}
      &                             &                         & Odd  & \textbf{1.336} & 1.325--1.347 & 8.477  & 6.473--11.10 & \phantom{$>$}0.86 \\
      \hline
    \end{tabularx}%
  \caption{\textbf{TTS for the exact solution, first-, and second-excited energy.} Estimated base value $b$ and constant $c$ controlling the exponential scaling in Eq.~\eqref{eq:scaling}, and associated $R^2$ values for the TTS linear fit across four algorithms (QAOA, Tabu Search, PCE, and PCE + Tabu) for even and odd instances. }
    \label{tab:fits_full_PCE}
\end{table*}

\subsection{Experimental deployment}\label{app:experiment}

\begin{figure}[!ht]
    \includegraphics[width=\columnwidth]{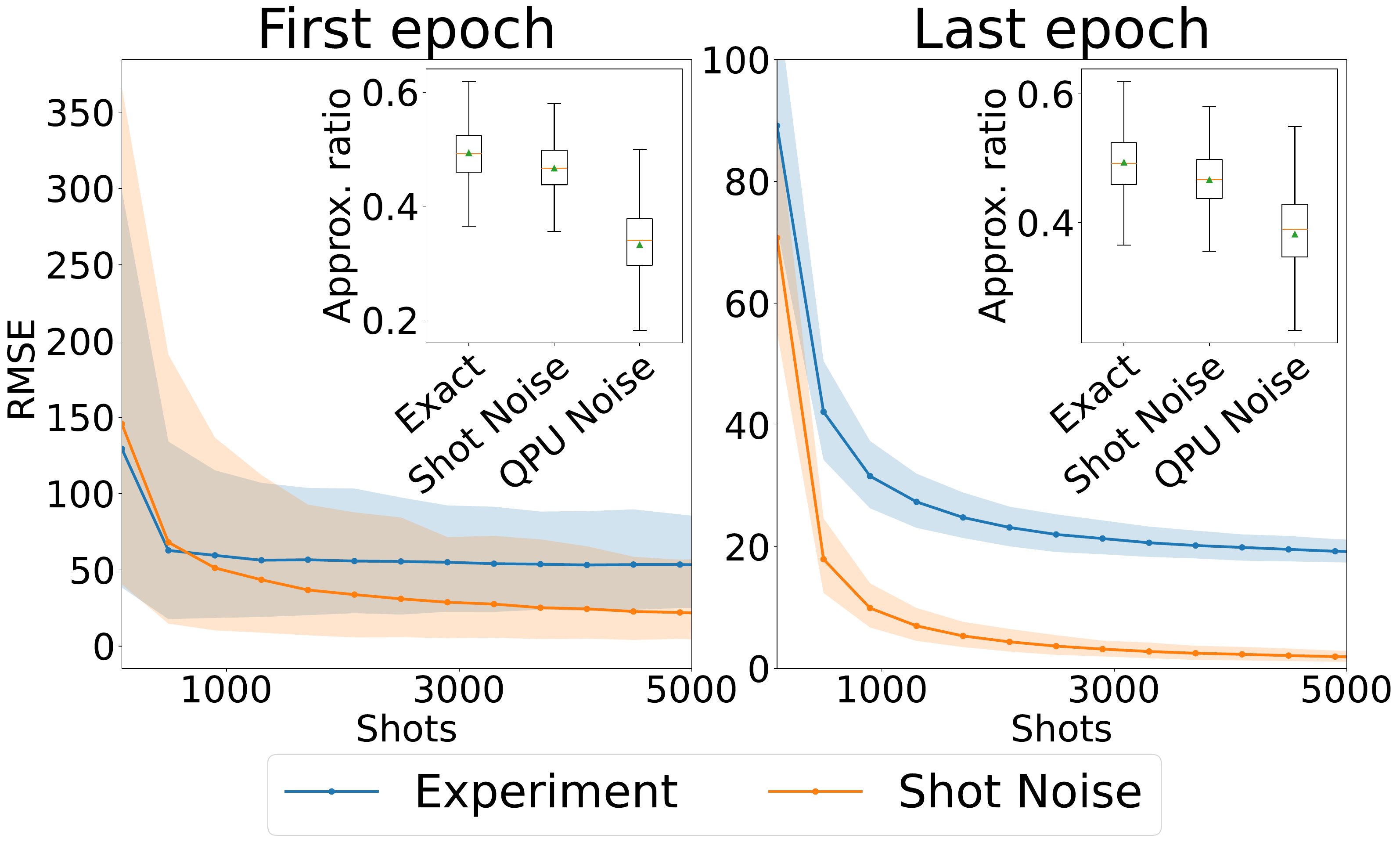}%
    \caption{\textbf{Experiment.} \textit{Main panels.} Relative mean-squared error (RMSE) of the measured loss values across three quantum states, evaluated at the first optimization epoch (left) and at the final epoch (right), as a function of the number of measurement shots. Experimental data were collected on the IonQ trapped-ion Forte-1 QPU. For comparison, we also report the corresponding shot-noise–only RMSE obtained from numerical simulations. The circuit uses 10 qubits and 30 two-qubit gates and targets a LABS instance of size $N=120$.
    \textit{Insets.} Box-plot distributions of the mean approximation ratio obtained over 500 random initializations by simulating the full optimization under three noise models: exact statevector simulation (Exact), shot-noise–only (Shot noise), and hardware-noise matching the observed QPU error rates (QPU noise). Each inset correspond to simulations using the RMSE evaluations in the main panels.  }
    \label{fig:experiment}
\end{figure}

We experimentally demonstrate our quantum solver on IonQ’s Forte-1 36-qubit trapped-ion processor using a LABS instance of size $N=120$. The device is based on trapped ytterbium ions and offers all-to-all qubit connectivity. We employ a standard VQA brickwork ansatz expressed in the hardware-native gate set, consisting of alternating layers of parametrized entangling 
RZZ gates and single-qubit rotation layers built from the native GPI and GPI2 operations. The circuit depth is fixed to 10 and includes 50 two-qubit gates.
We use the quadratic Pauli-correlation encoding and, for this experiment, restrict to the same-letter Pauli observables (XX, YY, ZZ) used originally in \cite{Sciorilli2025}, which reduces the measurement overhead to three bases. Circuit parameters are optimized offline via classical simulation, and the resulting pre-trained circuit is then executed on the hardware. For three independent random initializations of the ansatz parameters, we measure the loss function at the first and last optimization epochs.
For experimental run, we accumulated measurement statistics until the estimated values of the loss no longer changed.  Figure~\ref{fig:experiment} reports the relative error of the loss estimate as a function of the number of measurement shots.  
At the circuit depth that we tested, the effect of barren plateaus should be modest. Nevertheless, we observe a noticeable change in relative error between the first and the last optimization epoch. This effect is primarily due to the random initialization of the circuit parameters: at start, the state produces near-zero Pauli expectation values. In this regime, hardware noise and finite-shot fluctuations contribute a comparatively larger fraction of the measured value, thereby amplifying the relative error. As optimization progresses and the expectation values move away from zero, the signal-to-noise ratio improves and the relative error decreases accordingly.
Furthermore, we simulated the full optimization procedure assuming an average relative error on the loss function
equal to the one we observed experimentally. The resulting performance degradation is reported in the insets.
We observe that shot-noise alone causes only a marginal loss in solution quality, suggesting that a few thousand measurement shots already provide sufficient statistics for experimental training. In contrast, hardware noise does not prevent training, but it noticeably degrades the final solution quality, potentially eroding the expected scaling advantage in time-to-solution (TTS). On current QPUs, this can be addressed with polynomial-time quantum error-mitigation techniques. Alternatively, increasing the encoding compression rate could further reduce quantum resource requirements and exposure to noise, although this may introduce a trade-off with scaling performance. Overall, the results indicate a practical path forward: shot budgets are already manageable, but mitigating hardware noise is the crucial factor to preserve the observed advantage. 
Finally, in the last optimization epoch, in the rounding step that maps the relaxed experimental solution back to binary values, we observe a bitstring error rate of only $1.6\%$, indicating that the hardware error does not meaningfully affect the final solution quality. 
\begin{figure}[!ht]
    \includegraphics[width=\columnwidth]{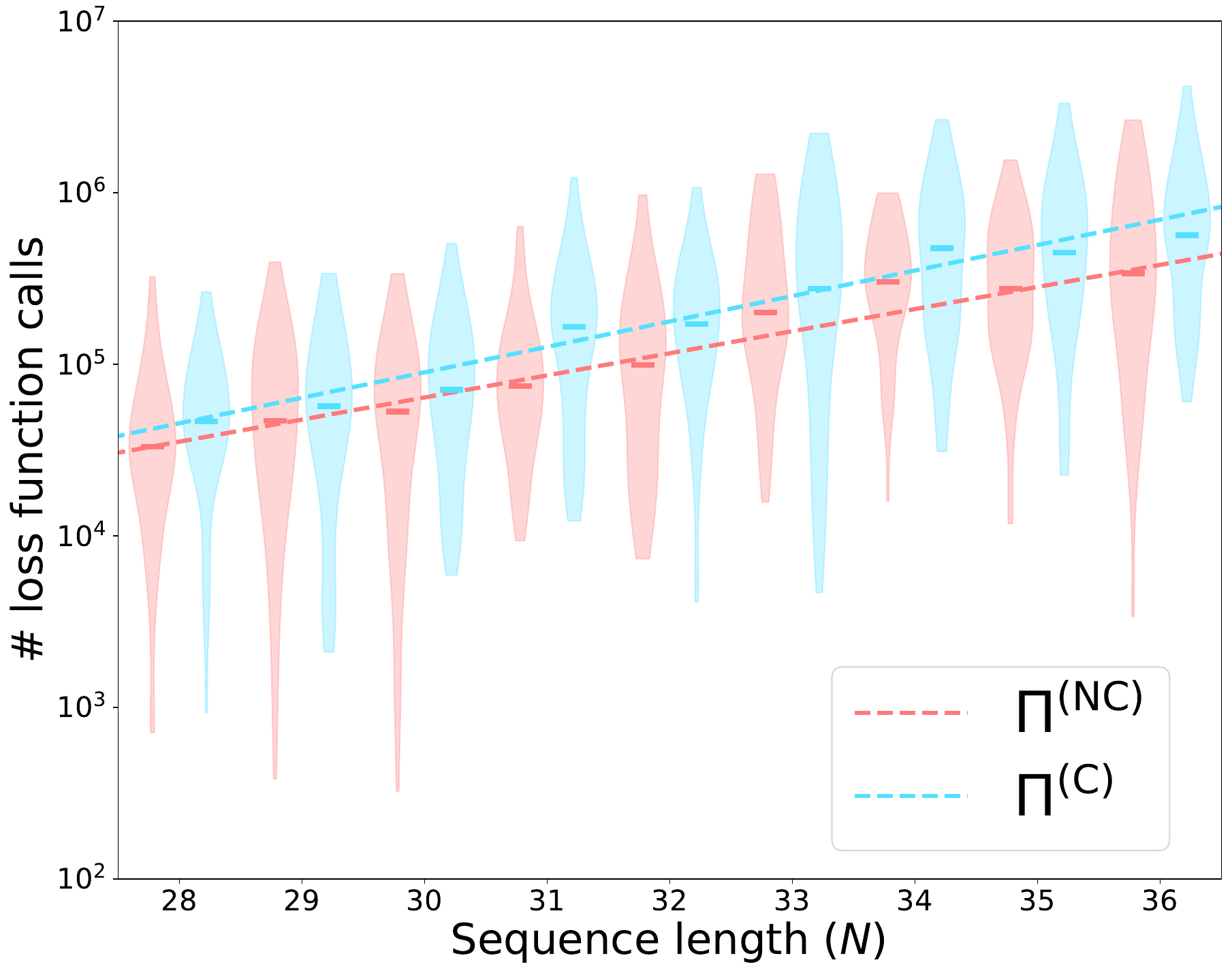}%
    \caption{\textbf{$\Pi^{(\text{C})}$ vs $\Pi^{(\text{NC})}$.} Scaling comparison of the $\text{TTS}$  for the quantum LABS solver based on Pauli Correlation Encoding using pauli sets chosen from $\Pi^{(\text{NC})}$  (red) \emph{vs.} pauli sets chosen $\Pi^{(\text{C})}$ ( blue) 
    for values of $N \in [28, 36]$. Here, TTS is defined as the total number of cost function evaluations required to reach the exact solution~\cite{BOSKOVIC2017262}. Dashed-dotted lines denote linear regression over the list of median values for each $N$. 
    The distribution of points over different initializations is depicted by violin plots at each $N$. The TTS distributions of  $\Pi^{(\text{C})}$ almost always dominate over the distributions of $\Pi^{(\text{NC})}$.
    Furthermore, the fits show a exponential scaling difference between the two approach, with a value of $b$ of $\order{1.345^N}$ for $\Pi^{(\text{NC})}$ against $\order{1.407^N}$ for $\Pi^{(\text{C})}$
 }
    \label{fig:min-max}
\end{figure}
\subsection{Building the Pauli sets $\Pi^{(\text{C})}$ and $\Pi^{(\text{NC})}$}
\label{app:noncommuting}

The two Pauli sets are randomly constructed as follows. 
First, we split the $4^n-1$ traceless Pauli strings in the Pauli group into subsets of maximally commuting operators using the procedure proposed in Ref.~\cite{bandyopadhyay2002}.  There are $2^n+1$ such sets, each having $2^n-1$ mutually commuting Paulis. 
Then, for each LABS instance size $N$ and for each optimization run, we repeatedly pick one such set and one Pauli string $P$ from it at random and accept or reject $P$ according to the following criteria: 
\begin{enumerate}[leftmargin=*]
    \item for $\Pi^{(\text{C})}$: accept $P$ if and only if it commutes with every other previously selected Pauli. 
    Repeating this eventually leads to a set of $N$ mutually commuting Paulis, as long as $N<2^n$;
    \item for $\Pi^{(\text{NC})}$: accept $P$ if and only if it anti-commutes with every other previously selected Pauli until reaching the maximum possible number $2n+1$~\cite{sarkar2019setscommutinganticommutingpaulis}). 
    After that, since it's impossible to anti-commute with all the others, accept those that anti-commute with the maximum possible number of selected Paulis. 
\end{enumerate}

In Fig.~\ref{fig:min-max}, we compare the performance of both sets and observed that $\Pi^{(\text{NC})}$ led to better TTS scaling and chose it as a figure of merit for the benchmark against other solvers.

\subsection{Comparison to State-of-the-Art Quantum Methods}\label{app:kike} 
For small instance sizes, the estimated time-to-solution (TTS) can vary a lot depending on which instance sizes are consider. To make a  fair comparison, we report our method’s performance on the same instances interval used for two state-of-the-art quantum methods Ref~\cite{cadavid2025} and Ref ~\cite{Shaydulin_2024}. We also compare against tabu search.
Furthermore, we investigate whether our solver can provide warm-start solutions to the memetic tabu search, following an analogous approach in Ref~\cite{cadavid2025}. Specifically, we take the best solution obtained across 150 independent PCE runs and copy it 50 times to initialize the population of the memetic tabu search. All hyperparameters of the classical algorithm are kept the same as in~\cite{cadavid2025}. For the time-to-solution (TTS) estimate, we include all loss-function queries used during the 150 PCE optimization runs, in addition to those performed by the memetic tabu search, all using the same bootstrap procedure detailed in~\cite{cadavid2025} to estimate the errors. 
The results are reported in Table ~\ref{tab:labs_results_L_ranges}

\begin{table*}[!t]
  \centering
  \begin{subtable}[t]{0.49\textwidth}
    \centering
    \begin{tabular}{|c|cccc|}
      \hline
      \textbf{Heuristic} & \textbf{Type} & \textbf{TTS} & \textbf{CI} & \textbf{$R^2$} \\
      \hline
      \multirow{2}{*}{Tabu Search} & Median   & 1.42 &1.26-1.60  & 0.83 \\
                               \cline{2-5}    & Ensemble & 1.45  & 1.41-1.49  & 0.43 \\
      \hline
      \multirow{2}{*}{PCE} & Median   &1.34  & 1.29-1.40 & 0.97 \\
                       \cline{2-5}     & Ensemble &1.33  &1.29-1.37 & 0.35 \\
      \hline
      \multirow{1}{*}{PCE + Mem. Tabu} & Median & 1.29 & 1.27-1.30  & 0.88 \\
      \hline
      \multirow{1}{*}{QE-MTS~\cite{cadavid2025} } & Median  &  1.24&1.23-1.25  &  0.88 \\
      \hline
    \end{tabular}
  \end{subtable}\hfill
  \begin{subtable}[t]{0.49\textwidth}
    \centering
    \begin{tabular}{|c|cccc|}
      \hline
      \textbf{Heuristic} & \textbf{Type} & \textbf{TTS} & \textbf{CI} & \textbf{$R^2$} \\
      \hline
      \multirow{2}{*}{Tabu Search} & Median   & 1.44 & 1.29-1.62 & 0.86 \\
                         \cline{2-5}         & Ensemble & 1.50 & 1.46-1.54 & 0.49 \\
      \hline
      \multirow{2}{*}{PCE} & Median   & 1.37 &1.32-1.43  &  0.97\\
                        \cline{2-5}    & Ensemble & 1.35  & 1.31-1.39  & 0.37 \\
      \hline
      \multirow{1}{*}{PCE + Mem. Tabu} & Median & 1.31 & 1.29-1.32  &  0.92\\
      \hline
      \multirow{1}{*}{QAOA ~\cite{Shaydulin_2024}} & Median  & 1.46 & 1.42-1.50 & 0.94 \\
      \hline
    \multirow{1}{*}{QAOA+QMF ~\cite{Shaydulin_2024}} & Median  & 1.21 & 1.19-1.23 & \textemdash \\
      \hline
    \end{tabular}
  \end{subtable}

  \caption{Estimated base value $b$, controlling the exponential scaling $\mathcal{O}(b^N)$, and associated $R^2$ values for the TTS linear fit across the considered heuristics, reported for two instance-length ranges : $ L \in [27, 38]$  on left and $L \in [28, 39]$ on the right.The two selected instance intervals match those used to benchmark the two state-of-the-art quantum methods.}
  \label{tab:labs_results_L_ranges}
\end{table*}

\subsection{Shot complexity}\label{app:sample}

Given $\varepsilon$, $\delta>0$, and a vector $\boldsymbol{\theta}$ of variational parameters, consider the problem of estimating $\calL$ up to additive precision $\varepsilon$ and with statistical confidence $1-\delta$. 
For each Pauli correlator $\expval{\Pi_i}$, $i\in[N]$, assume that, with confidence $1-\tilde{\delta}$, one has an unbiased estimator $\Pi_i^*$ with statistical error at most $\eta>0$, \emph{i.e.}, 
\begin{equation}\label{eq:est_error_Pi}
    \big|\Delta\langle \Pi_i \rangle\big| := \big|\langle \Pi_i \rangle - \Pi_i^* \big| \le \eta\,.
\end{equation}
The corresponding error in the loss function is $\Delta\calL := \calL-\calL^*$, with $\calL^*$ given by~\eq{loss} calculated using $\Pi_i^*$ instead of $\langle\Pi_i\rangle$. The multivariate Taylor theorem ensures that there is a $\boldsymbol{\xi}\in[-1,1]^{N}$ such that
\begin{align}
\label{eq:delta_calL}
\Delta\calL = \sum_{i\in[N]} \frac{\partial \calL}{\partial{u_i}}\bigg|_{u_i = \xi}\,\Delta\langle\Pi_i \rangle\,. 
\end{align}
For the loss function~\eq{loss}, using $\tanh(x)\le1$ and $\frac{d}{dx}\tanh(x)=\sech^2(x)\le1$, one can show that $\big|\frac{\partial \calL}{\partial u_{i}}\big|_{\xi{}_{i}} \le 2\alpha\big[N(N-1) + \beta\big]$. 
As a result, 
\begin{align}\label{eq:DeltaL}
|\Delta \calL|
&\le 2\,\eta\,\alpha\,N[N(N-1) + \beta]
\end{align}
follows immediately from the triangle inequality together with~\eq{est_error_Pi}. 
To ensure $|\Delta\calL|\le\varepsilon$ we then require that
\begin{align}\label{eq:etabound}
\eta \le \frac{\varepsilon}{2\,\alpha\,[N(N-1) + \beta]}\,.
\end{align}
The minimum number $S$ of samples needed to achieve such precision can be 
upper-bounded by standard arguments using the union bound and Hoeffding's inequality, which gives $S \le(2/\eta^2)\log(2 N/\delta)$.
Then, by virtue of Eq. \eq{etabound}, it suffices to take
\begin{align}
S \geq \frac{8\alpha^{2} N^{2}}{\varepsilon^2}\,[N(N - 1) + \beta]^{2}\log(\frac{2 N}{\delta}).
\end{align}
This is the general form of our upper bound and guarantees $P\left(|\Delta \mathcal{L}| \leq \varepsilon \right) \geq 1 - \delta$. 
Recall from App.~\ref{app:tuning} that $\alpha=1.5n=\mathcal{O}(\sqrt{N})$. 
We also reiterate that this a loose upper bound and, in practice, the actual number of samples is expected to have a better scaling in $N$.

\subsection{Quantum-classical crossover point}\label{app:crossover}
\begin{table}[b]

\renewcommand{\arraystretch}{1.3}
  \begin{tabularx}{.66\columnwidth}{@{\extracolsep{\fill}\hspace{0pt}}|l|c|c|c|}
    \hline
    \textbf{Solution} & \textbf{QPU} & \textbf{Adv.\ Type} & $\mathbf{N_*}$  \\
    \hline
    \multirow{4}{*}{Exact} &
      \multirow{2}{*}{SC} & Walltime & 1175  \\ \cline{3-4}
    &                      & Queries  & 1099     \\ \cline{2-4}
    & \multirow{2}{*}{TI} & Walltime & 1431  \\ \cline{3-4}
    &                      & Queries  & 1207   \\
    \hline
    \multirow{4}{*}{1st}  &
      \multirow{2}{*}{SC} & Walltime & 3705  \\ \cline{3-4}
    &                      & Queries  & 3469     \\ \cline{2-4}
    & \multirow{2}{*}{TI} & Walltime & 4407  \\ \cline{3-4}
    &                      & Queries  & 3763     \\
    \hline
    \multirow{4}{*}{2nd}  &
      \multirow{2}{*}{SC} & Walltime & 2597  \\ \cline{3-4}
    &                      & Queries  & 2433     \\ \cline{2-4}
    & \multirow{2}{*}{TI} & Walltime & 3107  \\ \cline{3-4}
    &                      & Queries  & 2645     \\
    \hline
  \end{tabularx}
  \caption{\textbf{Crossover points.} Estimated instance size $N_*$ after which the PCE solver is expected to outperform the classical Tabu search solver in terms of total walltime or total number of queries to the loss function for two different quantum hardware platforms: superconducting qubits (SC) and trapped ion (TI) qubits. 
  }
    \label{tab:crossover}
\end{table}

In our numerical results, we observed a competitive TTS scaling compared to Tabu Search in terms of query access to the loss function. 
However, each quantum query incurs an overhead given by gate operations and the sample complexity to guarantee that all expectation values are estimated within desired precision. Furthermore, our approach also need an estimation of the gradient, usually done in the context of VQA using parameter-shift-rule~\cite{Schuld2019}. 
Assuming the observed TTS scalings remain the same for larger instances, we can use the circuit and sample complexity estimates in Apps.~\ref{app:circuit} and~\ref{app:sample} to determine the crossover point in which we expect to observe a quantum advantage in query complexity. 
Our estimation in Tab.~\ref{tab:crossover} considers the scaling obtained for the the median of the even instances,as it has a better value of the $R^2$, and the crossing points for advantage are given in terms of query complexity and walltime for two different quantum hardware platforms: superconducting (SC) and trapped-ion (TI) qubits. For superconducting qubits, we take each layer to be executed in parallel, while, for trapped-ions, each gate is considered as executed sequentially. For the estimation of the walltime, we assumed an overhead of $5\times10^2$ for superconducting qubits~\cite{AbuGhanem2025}, and $6\times 10^6$ for trapped ion~\cite{ionq_aria_practical_performance_2025}, when comparing to a cycle of a current day CPU\cite{intel_i9_14900KS_2025}.
In addition, we disregarded any extra overhead due to error mitigation that the quantum approach may need.
This was done both because a detailed estimation would be heavily platform dependent, and because the actual impact that error noise have in the training of our specific VQA (and so the level of error mitigation actually needed) is still an open question. 
While the instance sizes in Tab.~\ref{tab:crossover} indicate the expected scaling crossover between the two methods, PCE also shows gains in producing high-quality approximate solutions. This suggests that PCE may outperform the classical baseline at substantially smaller sizes, especially given Tabu Search’s sharp performance degradation for instances of a few hundred variables. Furthermore, as also observed in~\cite{Sciorilli2025}, our shot-complexity estimate is an upper bound and is likely conservative. In practice, we expect fewer circuit executions to suffice, which would lower the crossover point accordingly.

Finally, the algorithm relies on many circuit re-initializations but requires only a modest number of qubits. This makes it naturally suited to massively parallel execution across many small QPUs, potentially further reducing threshold needed to observe a walltime advantage.

\end{document}